\documentclass[twocolumn,showpacs,preprintnumbers,amsmath,amssymb]{revtex4}

\usepackage{graphicx}
\usepackage{dcolumn}
\usepackage{bm}

\newcommand{\ket}[1]{\ensuremath{\left|#1\right\rangle}}

\begin{document}

\preprint{1}

\title{Stimulated generation of superluminal light pulses via four-wave mixing}

\author{Ryan T. Glasser\footnote{R. Glasser and U. Vogl contributed equally to this work.}}

\author{Ulrich Vogl\footnotemark[\value{footnote}]}
\author{Paul D. Lett}
\affiliation{
National Institute of Standards and Technology \\and Joint Quantum Institute, \nolinebreak NIST and University of Maryland, Gaithersburg, MD 20899 USA\\
}

\date{\today}

\begin{abstract}
We report on the four-wave mixing of superluminal pulses, in which both the injected and generated pulses involved in the process propagate with negative group velocities.  Generated pulses with negative group velocities of up to $v_{g}=-\frac{1}{880}c$ are demonstrated, corresponding to the generated pulse's peak exiting the 1.7\,cm long medium $\approx50$\,ns earlier than if it had propagated at the speed of light in vacuum, $c$.  We also show that in some cases the seeded pulse may propagate with a group velocity larger than $c$, and that the generated conjugate pulse peak may exit the medium even earlier than the amplified seed pulse peak.  We can control the group velocities of the two pulses by changing the seed detuning and the input seed power.  

\end{abstract}


\maketitle

 There has been substantial recent interest in modifying the group velocity of optical pulses, resulting in slow, stopped, and superluminal light \cite{2010review}.  Experiments have resulted in extremely slow light with very small group velocities \cite{zac2004,gaeta2005,gauthier2007,hau1999,camacho2006,boyer2007,camacho2009}, while others exhibit the ability to store and retrieve optical pulses \cite{zac2001,manson2005,hemmer2000,lukin2000}.  It is also possible to generate a dispersion relation that results in negative group velocities \cite{garrett1970,macke1985,chiao1994,bolda1994,dogariu2000,gisin2004,milonni2002,boyd2003,thevenaz2005,Zhang2006,Pati2009,Kuang2009,Luo2010}.  In such cases when the group velocity of light in a material is negative, the exiting pulse's peak can appear to exit the medium before the peak of the input pulse enters.  The peak that exits the medium does not correspond point-to-point to the peak of the input pulse.  Rather, the phenomena of re-phasing allows for different frequency components of the input pulse to propagate at different velocities, which results in the apparent pulse peak advancement seen in fast light experiments \cite{macke2003,wang2001}.

 We report the stimulated generation of light pulses that propagate with a group velocity faster than the speed of light in vacuum, via four-wave mixing (4WM) in hot rubidium vapor.  The 4WM process employed here involves injecting one weak beam into the medium and pumping with a beam at a different frequency, as seen in Fig. 1.  A beam at a third frequency is generated via the process, as photons from the pump beam are converted into photons in the injected seed and generated conjugate modes.  The amplified seed pulse is shown to have a negative group velocity due to the 4WM dispersion, and stimulates the generation of the conjugate pulse that may appear to propagate even faster, as seen in Fig. 2.  The anomalous dispersion results from asymmetric gain and absorption lines at the seed and generated conjugate pulse frequencies.  We show that it is possible to manipulate the group velocities of the two modes relative to one another, to some extent, by detuning the seed pulse or varying the input seed intensity.  The scheme could be applicable to a variety of optical communications scenarios, where the correction of pulse jitter by advancing or delaying pulses may be necessary.  The present results will allow us to investigate the effects of superluminal group velocities on quantum entanglement and squeezed light, both of which may be produced via the 4WM process \cite{boyer2007,lett2007}.

The first experiments to produce fast light used absorption lines \cite{wong1982,macke1985,garrett1970}, which exhibit strong negative dispersion at the center of the line, but the dispersion is accompanied by significant attenuation of the input pulses.  More recently, schemes using gain doublets have been shown to produce fast light (\cite{2010review,chiao1994,dogariu2000} and references therein).  All previous experiments involving fast light, to our knowledge, involve injecting a pulse into the fast light medium, and observing that it appears to exit the medium faster than a reference pulse traveling at the speed of light in vacuum.  In our experiment, we show not only that the injected pulse propagates with a negative group velocity, but also that a second pulse at another frequency and in a separate spatial mode is generated by the 4WM process, which may appear to propagate faster than a vacuum-traversing reference pulse as well. In contrast to some previous 4WM experiments \cite{boyer2007}, the group velocities of the seed and conjugate modes under the fast light conditions in the present experiment exhibit little coupling due to the large differential absorption between the two modes.  The gain line for the generated conjugate frequency sits inside the edge of an absorption line, resulting in conjugate pulses with a significantly weaker amplitude than the amplified seed pulses which are far off-resonance.  Previous theoretical investigations involving four-level atomic systems to generate fast light have been performed in which the probe field is co-propagating with the coupling field \cite{Kuang2009}.  Additional theoretical work has shown that the double-lambda scheme and electromagnetically induced transparency may result in slow and fast light for two input probe beams \cite{Patnaik2011}.  Experimental work has been done in which Raman gain of an input probe results in slow light, while pump depletion results in anomalous dispersion at the pump frequency (heterodyne detection was used to resolve the different frequency components) \cite{Pati2009}.  The present experiment differs from these theoretical and experimental investigations in that we generate a conjugate pulse in a separate spatial mode that may propagate superluminally, in addition to the seed pulse.  This also has the benefit of allowing for direct detection of both pulses, rather than having to use heterodyne detection to resolve the separate frequency modes.

For the generation of a strong anomalous dispersion we use the 4WM process shown schematically in Fig. 1.   The gain line that the input seed pulse experiences results in negative dispersion at the gain line edges, giving rise to superluminal propagation of the injected pulse.  The generated conjugate pulse experiences a broad region of negative dispersion due to the gain and absorption line profile resulting from this 4WM process, which is asymmetric (see Fig. 1) and exhibits absorption on the high frequency side of the gain line \cite{boyer2007,lett2007}.  Here the seed and conjugate pulses sample independent dispersion features approximately 6\,GHz apart, in contrast to the gain-doublet scheme that is sometimes used to create a single dispersion feature \cite{chiao1994}.

\begin{figure}
\includegraphics[height=5.9cm]{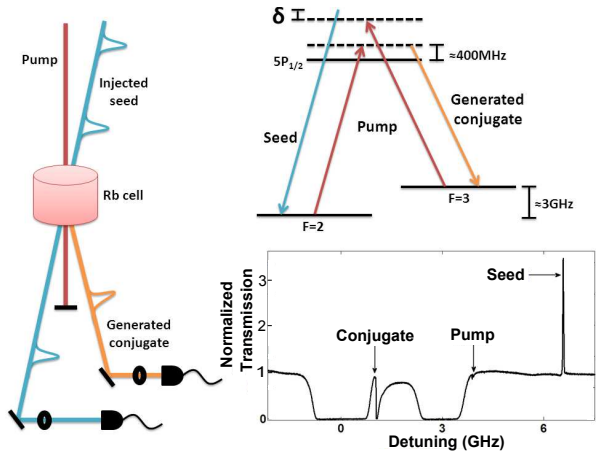}
\caption{\label{fig:epsart} Experimental setup and double-lambda scheme.  The pump laser is detuned approximately 400\,MHz to the blue of the Rb D1 line.  Generated conjugate pulses are created inside the cell, and exit at the same angle as the seed pulses.  The generated conjugate is on the wing of an absorption line, while the injected seed is blue detuned $\approx$\,6\,GHz relative to the conjugate.  The double-lambda scheme is shown, and the two-photon detuning $\delta$ is indicated.  The lineshapes of the gain lines at the seed and conjugate frequencies are also shown.}
\end{figure}

The 4WM process involves the annihilation of two pump photons, and the creation of a single probe and conjugate photon, as is evident when examining the simplified phenomenological interaction Hamiltonian,
\begin{equation}\label{1}
\hat{H}_{I}=\chi\hat{a}^{\dag2}\hat{b}\hat{c}+\chi^{*}\hat{a}^{2}\hat{b}^{\dag}\hat{c}^{\dag}
\end{equation}
\cite{yuen1976,walls1986}.  Here, the modes $\hat{a}$, $\hat{b}$ and $\hat{c}$ correspond to the pump, injected seed and generated conjugate, respectively, and $\chi$ is the effective interaction strength that depends on the third-order susceptibility and the length of the interaction.  Combined with energy conservation, this results in the constraint on frequencies such that $2\omega_{pump}=\omega_{s}+\omega_{c}$.  The generated conjugate pulses in the present experiment experience both absorption and gain, whereas the seed pulses experience only gain, as seen in Fig. 1.  

\begin{figure}
\includegraphics[height=6.0cm]{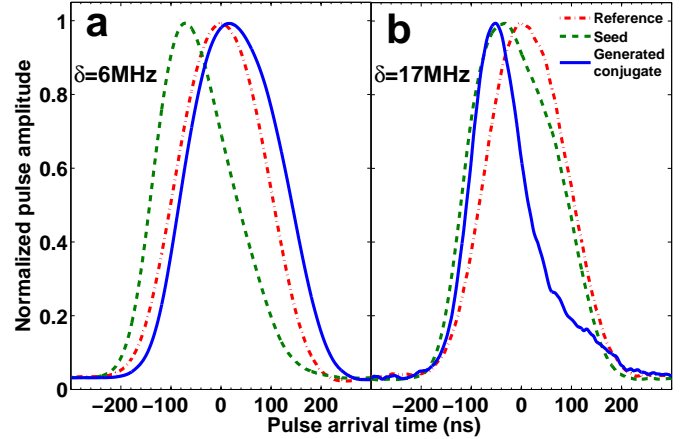}
\caption{\label{fig:wide}  Traces of the amplified seed and generated conjugate pulses exhibiting negative group velocities, for two different seed pulse detunings.  The blue, green dashed and red dot-dashed curves correspond to the generated, amplified seed and reference pulses, respectively.  The pulse amplitudes have been normalized relative to the reference pulses, and are averaged 512 times.  The amplified seed and generated pulses on the left are scaled by a factor of 0.36$\times$ and 0.32$\times$ relative to the reference pulses.  On the right figure, the seed pulse is scaled by a factor of 0.12$\times$ and the generated pulse is scaled by 0.78$\times$ relative to the reference.}
\end{figure}

The 4WM process is pumped with a strong ($\approx$\,220\,mW) continuous-wave linearly polarized laser detuned $\approx\,$400\,MHz to the blue of the Rb D1 line $\ket{5S_{1/2},F=2}\rightarrow\ket{5P_{1/2}}$, at $\lambda\,\approx\,795$\,nm (see Fig. 1).  Weak input seed pulses with peak powers of $\approx\,$5$\,\mu$W with a full-width at half-maximum (FWHM) of $200$\,ns and a frequency bandwidth of $\approx5$\,MHz, are injected at an angle of $\approx1^{\circ}$ relative to the pump.  The seed pulses are orthogonally polarized and detuned $\approx3$\,GHz to the blue of the pump beam.  The injected seed pulse frequency is varied by changing the operating frequency of a double-passed 1.5\,GHz acousto-optic modulator, which is used to generate the seed from part of the pump beam.  This allows us to precisely control the relative frequency detuning between the pump and seed beams, though the absolute position of the pump beam frequency varies slightly in each experimental run.  The pump and probe are focused into the center of the cell with focal spot sizes of $\approx800\,\mu m\times1000\,\mu$m and $\approx700\,\mu m\times700\,\mu$m, respectively.  A conjugate pulse is created which propagates at the same angle as the seed relative to the pump, but in the opposite azimuthal direction.  The $^{85}$Rb cell is 1.7\,cm long and is held at a temperature of $\approx116^{\circ}$C.  The seed and generated beams are spatially filtered with irises after the cell in order to select only the central spots of the 4WM beams and filter out residual pump light.  Approximately $2/3$ of the pulse power is detected.  The pulses are then detected with a high-gain avalanche photodiode operating in the linear mode and fed directly to an oscilloscope and averaged 512 times.  Reference pulses are obtained by measuring the seed pulses when the pump beam is blocked, and serve as a measure for the vacuum propagation speed of the pulses for each experimental run.  The reference pulses were also measured with the Rb cell removed, to ensure that they propagate at the same speed with and without the cell present (when the pump is blocked) to within our experimental time resolution.

Most previous experiments involving fast light use absorption lines, single gain lines, or double gain lines that are Lorentzian in shape to generate the desired dispersion \cite{2010review,garrett1970,chiao1994,dogariu2000}.  In the present experiment, the generated conjugate experiences an asymmetric gain line with an absorption dip on one of the wings, while the seed experiences only a gain line, as seen in Fig. 1 \cite{boyer2007}.  Modeling the line shape for the generated conjugate mode as a Lorentzian for the gain and a second Lorentzian for the absorption results in:
\begin{align*}
k(\omega)=&\frac{\omega}{c}n_{0}+\tag{2}\\
&\frac{1}{2}\left(\frac{\alpha_{g}\gamma_{g}}{(\omega-\omega_{g})+i\gamma_{g}}+\frac{\alpha_{a}\gamma_{a}}{(\omega-\omega_{a})+i\gamma_{a}}\right)
\end{align*}
Here $n_{0}$ is the background index, and $\alpha_{g}$ and $\alpha_{a}$ are the coefficients of the gain and absorption components, respectively, with $\alpha_{g}<0$ resulting in gain. The operating frequency and the center frequencies of the gain and absorption lines are denoted by $\omega$, $\omega_{g}$, and $\omega_{a}$.  The linewidths of the gain and absorption lines are $\gamma_{g}$ and $\gamma_{a}$, respectively.  Due to the large absorption present at the generated pulse frequency, we use the maximum gain of approximately $G\,=\,e^{-\alpha_{g}L}=20$ measured at the seed frequency to determine the gain coefficient in the presence of absorption.  The modeled gain line is then fit to the measured gain line, which has a FWHM of $\approx$\,20\,MHz, as well as an absorption dip at the wing.  The parameters in the model are $\alpha_{g}=-175$\,m$^{-1}$, $\gamma_{g}=20$\,MHz, $\alpha_{a}=95$\,m$^{-1}$ and $\gamma_{a}=23$\,MHz.

The group velocity $v_{g}$ is defined by
\begin{align*}
k_{1}=&\frac{dk}{d\omega}=\frac{1}{v_{g}}= \tag{3}\\&\frac{n_{0}}{c}-\frac{\alpha_{g}\gamma_{g}}{2(\omega-\omega_{g}+i\gamma_{g})^{2}}-\frac{\alpha_{a}\gamma_{a}}{2(\omega-\omega_{a}+i\gamma_{a})^{2}},
\end{align*}
where $k_{1}$ is the first term in the power series expansion of $k(\omega)$ about the central frequency.  The resulting expected time delays and advancements from this model are shown in Fig. 3, with the inset showing a comparison between the measured and modeled conjugate lineshape.  Both the magnitude and detuning dependence of the pulse advancement/delay fit the experiment remarkably well.  The seed gain line is modeled as a single gain lineshape that has an asymmetric lineshape as measured (modeled as the sum of several Lorentzians to account for the asymmetry).  We speculate that the modeled relative advancements for the seed pulses do not fit the measured data very well due to the steep dispersion of the seed gain line.  This results in the seed pulse's bandwidth exceeding some regions of ``linear" dispersion, making higher order terms in the power series expansion of $k(\omega)$ contribute more significantly.
\begin{figure}
\includegraphics[height=6.0cm]{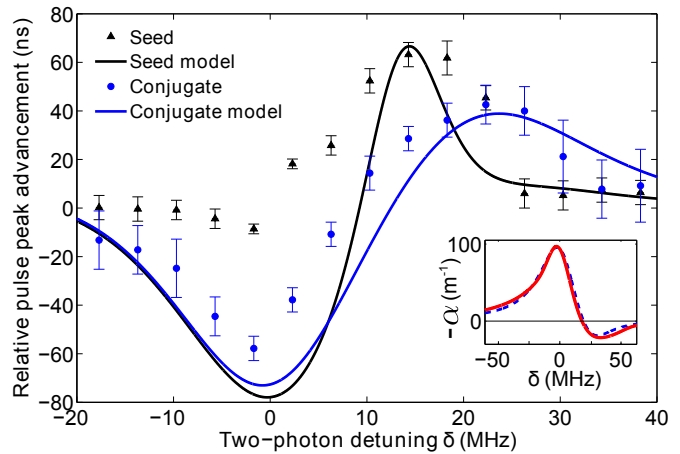}
\caption{\label{fig:epsart} Generated conjugate and amplified seed pulse peak advancement versus relative seed detuning.  The black triangles correspond to the amplified seed pulse peak advancement when the pump is present. The blue circles correspond to the generated conjugate pulse peak advancement.  Below seed detunings of $\approx-20$\,MHz, the generated conjugate is too weak to be measured.  The input seed peak power is approximately $5\,\mu$W.  The blue curve is the generated pulse's expected advancement derived from the double-Lorentzian gain and absorption line model corresponding to the solid curve in the inset. The inset shows the frequency dependent total absorption coefficient $\alpha=$Im$(n(\omega))$, with $n(\omega)=\frac{ck(\omega)}{\omega}$ (plotted is -$\alpha$, so that gain corresponds to positive values). The dotted curve in the inset is the measured conjugate lineshape.  The black curve is the expected advancement for the modeled seed gain line.}
\end{figure}
Experimentally the group velocity is determined by the measured arrival time of the pulse peak relative to a reference pulse propagating at $c$.  The next order term is the lowest-order term that contributes to pulse reshaping, as it describes dispersion in the group velocity.  The superluminal generated pulses in this experiment exhibit some reshaping relative to the reference pulses.  The amount of pulse reshaping is sensitive to various experimental parameters such as beam waists, cell temperature and input pulse width.  In the present experiment we operated with parameters that allow for minimal reshaping, but consequently sub-optimal pulse advancement.

When the strong pump beam is present, it is possible for both the injected seed and generated conjugate pulses to exhibit negative group velocities, depending on the pump and injected seed detunings.  A negative group velocity can be understood by considering its relation to a pulse's arrival time delay, $\Delta T=\frac{L}{v_{g}}-\frac{L}{c}$, after propagating some distance $L$.  When the group velocity is negative, so is the time delay, which corresponds to the pulse peak exiting the medium at a time $-\Delta T$ sooner than if a similar pulse had traversed the same distance in vacuum.   With a cell length of 1.7\,cm, we obtain a maximum pulse peak advancement of 50\,ns for the generated pulse, which corresponds to a group velocity of $v_{g}=-\frac{1}{880}c$, and a relative pulse peak advancement of $25\%$.  This is the first observation of stimulated generation of superluminal pulses, in particular via the 4WM interaction with an injected seed. The peak of the stimulated conjugate pulse exits the medium significantly before the peak of the seed pulse enters.  In vacuum, the seed pulse would traverse the Rb cell length in $\approx$\,0.057\,ns, but the generated conjugate's peak exits $\approx$\,50\,ns earlier when optimized for minimal pulse distortion.  Allowing for more severe distortion, we have obtained a maximum conjugate pulse peak relative advancement of 90\,ns.

In Figs. 2b and 3 we show cases where the amplified seed pulse exhibits a negative group velocity and the generated conjugate pulse's group velocity may be tuned such that the pulse peak exits the medium prior to the exit of the peak of the seed pulse (all uncertainties shown in the figures are one standard deviation, combined statistical and systematic uncertainties).  As previously mentioned, the absolute frequency of the pump beam varies somewhat during each experimental run.  This has the effect of slightly changing the position of the gain lines as well their shape.  The relative pulse peak advancement as a function of the two-photon detuning maintains the same overall shape as in Fig. 3 despite this.  One consequence is that there may be slight shifts in the relative advancements between the seed and conjugate pulses.  This results in the conjugate pulse in Fig. 2b being slightly more advanced than the seed pulse when the two-photon detuning is set to 17\,MHz, a variance from what is indicated in Fig.\,3.  The generated pulse is somewhat reshaped relative to the seed pulse, but the pulse peak and leading edge at half-maximum are significantly advanced relative to the reference pulse.  At a detuning set for the maximum advancement of the generated pulse relative to the injected pulse ($\delta\approx$23\,MHz), the generated pulse has approximately $20\%$ of the peak amplitude of the reference (input) pulse.  Additionally, under the same conditions, the generated conjugate pulse is shown to exit the superluminal medium with a pulse peak advancement of 8$\%$ relative to the superluminal seeded pulse, and 25$\%$ relative to a 200\,ns FWHM reference pulse traveling at $c$.  We note that this advancement is achieved for a peak gain of $G\,\approx$\,20, which is a relatively large advancement considering it has been theoretically shown that for systems consisting of an absorption line in the center of a broad gain background of $G\,=\,e^{32}$ the max relative advancement is only $2\sqrt{2}$ \cite{Narum2010}.

\begin{figure}[t]
\includegraphics[height=6.2cm]{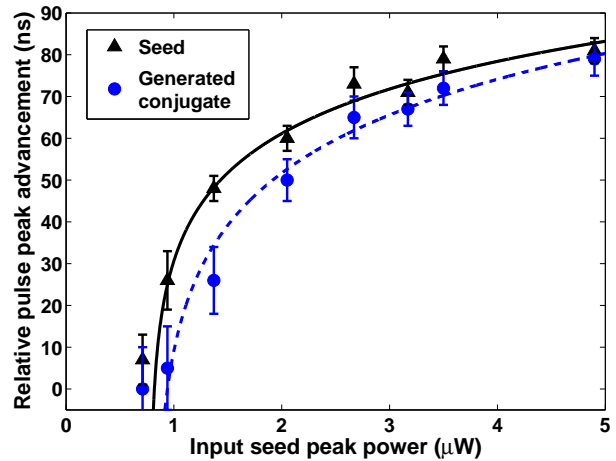}
\caption{\label{fig:epsart} Generated conjugate and amplified seed pulse peak advancements versus normalized input seed power at a relative detuning of $\delta\approx18$MHz.  The black triangles correspond to the amplified seed pulse and the blue dots correspond to the generated conjugate pulse.  The curves are fits to the data as a function of the logarithm of the input seed power.}
\end{figure}

The measured pulse peak advancement for both the amplified seed pulse and the generated conjugate pulse as a function of the seed pulse detuning is shown in Fig. 3.  For a variety of detunings the generated pulse closely resembles the input seed pulse shape.  We have some flexibility in controlling the relative group velocities of the amplified seed and generated conjugate pulses, in particular we see cases when both pulses are superluminal, with the generated pulse peak exiting the medium first.  At the largest detunings the generated pulse exhibits significant reshaping relative to the seed.

The generated pulse and amplified seed pulse advancements versus injected seed power are shown in Fig. 4.  This situation is unusual in that the intensity of the beam influences its own group velocity.  This is somewhat analogous to a Kerr medium in which the intensity-dependent index of refraction alters the phase velocity.  To account for the nonlinear, intensity-dependent part of the refractive index, typically the nonlinear Schr\"{o}dinger equation is used \cite{Lisak1983}.  We speculate that in addition to this, group velocity dispersion and higher order terms, as well as some coupling between the seed and generated beams all contribute to the nontrivial intensity dependence of the relative pulse peak advancement.  Despite this complexity, the pulse advancement is well approximated by a natural logarithm dependence on the input intensity, as seen by the fits in Fig.\,4.  The generated conjugate pulse advancement is also controllable by varying the input seed power, resulting in a tunable advancement.

We have shown that a superluminal seed pulse can stimulate the creation of an additional superluminal conjugate pulse by the 4WM interaction.  This generated pulse can propagate even faster than the superluminally-propagating seed pulse, with relatively small distortion.  The pulse peak advancements of the two pulses are tunable by changing the input seed detuning and power.  This could have applications in optical communication schemes in which pulse jitter may be compensated for by advancing or delaying pulses accordingly.  In an on-off keying optical communications system, if the jitter is small relative to the pulse width, it may be beneficial to advance a pulse rather than delay the entire pulse train.  The benefits of applying this type of jitter correction would have to be weighed against the disadvantages, which include added noise and some degree of pulse reshaping.  Additionally, due to the multi-spatial-mode nature of 4WM in atomic vapors, the present results suggest that the superluminal propagation of images may be possible in future experiments.  The high level of squeezing obtainable via 4WM, and the fact that fast light may also be obtained at a relatively low gain (and added noise), suggests that one may be able to use the quantum correlations between the twin beams to further investigate the details of superluminal light pulse propagation in such media.  In particular, one can hope to use the quantum correlations to examine experimentally which part of the input pulse is causally linked to the peak of the output pulse.

\begin{acknowledgments}
This work was supported by the Air Force Office of Scientific Research.  This research was performed while Ryan Glasser held a National Research Council Research Associateship Award at NIST.  Ulrich Vogl would like to thank the Alexander von Humboldt Foundation.

\end{acknowledgments}


\end{document}